# A Review of Noise Cancellation Techniques for Cognitive Radio

ADNAN QUADRI
Department of Electrical Engineering
University of North Dakota
Grand Forks, ND, 58203
USA
adnan.quadri@und.edu

*Abstract:* - One of the fundamental challenges affecting the performance of communication systems is the undesired impact of noise on a signal. Noise distorts the signal and originates due to several sources including, system non-linearity and noise interference from adjacent environment. Conventional communication systems use filters to cancel noise in a received signal. In the case of cognitive radio systems, denoising a signal is important during the spectrum sensing period, and also during communication with other network nodes. Based on our findings, few surveys are found that only review particular denoising techniques employed for the spectrum sensing phase of cognitive radio communication. This paper aims to provide a collective review of denoising techniques that can be applied to a cognitive radio system during all the phases of cognitive communication and discusses several works where the denoising techniques are employed. To establish comprehensive overview, a performance comparison of the discussed denoising techniques are also provided.

## 1 Introduction

Noise cancellation is one of the main challenges in communication systems and has been a focus of study for many years. Techniques and technologies for noise cancellation have emerged from the need to mitigate unwanted noise present in the desired signal. Noise distorts the received signal in a random manner and occurs because of several sources. According to [1–12] some of the prominent sources are a) non-linearity present in the RF front-end, b) time-varying thermal noise at the receiver of a communication system, and c) noise interference from adjacent environment. In addition, there are several other factors affecting the received signal such as crosstalk and electromagnetic interference.

Over the past few decades, several denoising techniques have been proposed [12-15]. These techniques can be classified into two categories: gradient-descent and non-gradient based adaptive filter algorithms. Gradient descent also referred to as steepest descent, are multivariate optimization techniques that start with an assigned initial value and follows the negative of the gradient to reach the desired local minimum. Examples of gradient-descent based adaptive algorithms are Least-Mean-Square (LMS) and its variants. Non-gradient algorithms include evolutionary algorithms based noise cancellation.

However, in the context of cognitive radio (CR) systems [16], few research papers on noise cancellation has been published, which might be because the cognitive radio technology itself is an emerging communication technology. Conventional communication systems use filters to cancel noise during communication. Besides noise cancellation during usual communication, a CR system can also employ denoising techniques during the spectrum sensing phase to increase the accuracy of sensing [9-15]. Although few survey papers are found that review denoising techniques for the spectrum sensing phase of cognitive communication, collective review of denoising techniques applicable to all the communication phases of cognitive radio have not been published yet. This paper aims to provide an overview of denoising techniques that are applicable to all the communication phases of the cognitive radio network and give a performance analysis of these techniques. As shown in Figure 1, denoising techniques can be classified into three categories: 1) Time-frequency analysis, 2) Matrix factorization, and 3) Adaptive filter based techniques.

Time-frequency analysis based denoising techniques allow inspection of the noise-induced signal in both time and frequency domain. Examples of techniques under this category are empirical mode decomposition and wavelet-based denoising. Although based on the same method of analysis, the approaches for denoising a signal is different for the two time-frequency analysis based techniques.

The second category, matrix factorization based denoising techniques, provides the means to perform signal space analysis. Examples of matrix factorization techniques are singular value decomposition and non-negative matrix factorization based denoising. Both singular value decomposition and non-negative factorization are capable of factorizing a huge or sparse matrix into smaller data sets, which allows easier inspection of a signal.

The third category, adaptive filter based denoising techniques, performs noise cancellation by employing filters with adaptive algorithms. Least-Mean-Square (LMS) and Normalized LMS based adaptive filters are examples of techniques under this category. LMS and NLMS based adaptive filters are capable of readjusting their filter parameters to cancel noise from a signal.

The rest of the paper is organized as follows: denoising techniques and the three categories are described in section II. In section III, this paper provides a performance analysis and comparison of the denoising techniques in terms of the performance criterion, strengths, and weaknesses. Finally, in section IV, conclusions are drawn.

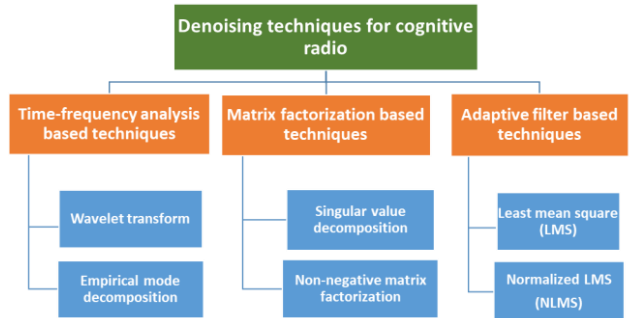

Fig. 1. Classification of noise cancellation techniques for cognitive radio

## 2 Classification of Denoising Techniques for Cognitive Radio

As previously explained, denoising techniques are classified into three categories, as shown in Figure 1. Each technique under the three categories is discussed and described in this section.

## 2.1 Time-Frequency Analysis Based Techniques

Time-frequency analysis allows inspection of a signal in both time and frequency domain. Such analysis is of great significance in case of detecting changes or singularities in the spectral content [16]. Examples of techniques under this category are wavelet transform and empirical mode decomposition.

### 2.1.1 Wavelet Transform Based Denoising

Conceptualized since the late 1980's, wavelet transform allows signal processing for time-frequency analysis. Succinctly put, a mother wavelet is chosen, which is also referred to as a basis function and is the primary step to wavelet analysis of a received signal. Wavelets revolve around the basis function by using a shifted and dilated version of the function. Translation and dilation introduce enough components to the transformation to retain the main properties of the original signal. In [16], authors put forward two important properties of wavelets which are admissibility and regularity conditions. The first property, admissibility allows decomposition of a signal which can later be reconstructed without losing any of the components of the original signal. Such breakdown enables identification of noise which is spread over a large number of coefficients, unlike the main signal which usually is found in a small portion of wavelet dimensions [17 - 18]. More elaborate discussion on the wavelet properties can be found in [19 - 20].

Two type of wavelet transform is continuous wavelet transform (CWT) and discrete wavelet transform (DWT). In paper [21], authors define the CWT for a signal x(t) as the sum of time components of the signal multiplied by the scale and shifted versions of the wavelet $\emptyset(t)$, which can be written as –

$$W(a,b) = (1/\sqrt{a}) \int_{-\infty}^{\infty} x(t)\psi((t-b)/a)\, dt, \quad (1)$$

Where, $a$ is the scaling factor, $b$ is the shifting factor, $x(t)$ is the signal and $\psi((t-b)/a)$ is the shifted and scaled wavelet $\emptyset(t)$. Reconstruction of the signal after analysis can be achieved by doing an inverse of the transform, as in (2) –

$$x(t) = (1/C_\psi) \int_{-\infty}^{\infty} W(a,b)\, \psi((t-b)/a)\, da\, (db/|a^2|) \quad (2)$$

Where,

$$C_\psi = \int_{-\infty}^{\infty} (|\varphi(\omega)|^2)/|\omega|\, d\omega, \varphi(\omega) \quad (3)$$

being the fourier transform of $\emptyset(t)$. Discrete wavelet transform (DWT) which is the sampled version of continuous wavelet transform (CWT), generates successive low-pass and high-pass filters of the discrete time-domain signal. Like the scaling and shifting factor of CWT, high-pass filter produces detailed information $d(n)$ and low-pass filter produces approximations $a(n)$ at each level of DWT. Two approaches to denoising with DWT are 1D wavelet denoising and 2D wavelet denoising. 2D wavelet denoising are well known for use in image compression and denoising [22].

In [8], a two-branch wavelet-based denoising is proposed, a technique which aims to locate noise singularities for the purpose of denoising and be able to reconstruct the original signal. Two-branch wavelet denoising goes through two stage of denoising in the first branch and the second branch is initiated only when it is found to be necessary thus reducing computational redundancy. During the first stage, Lipschitz exponent and wavelet transform modulus maxima is used for denoising. Details of the mentioned stage involve locating noise singularities at each scale to eventually remove the found modulus maxima. In [23], one of the methods for edge detection is based on wavelet transform modulus maxima, which is effective in locating singularities under high signal-to-noise ratio (SNR) and the other method is based on Multiscale wavelet product which enhances the multiscale peaks due to the edges and makes it convenient to detect noise-induced singularities.

Several other application of wavelet-based noise mitigation may not have been introduced to cognitive radio technology yet but sparsely refers to the possibility of being effective once employed. For instance, in [24] wavelet-based denoising technique is tried to get better power delay profile estimates in indoor wideband environments. Work in [25] presents an estimation of TDOA – time difference of arrival for GSM signals in noisy channels using wavelet-based denoising technique. While [26] discusses composite wavelet shrinkage for the purpose of denoising low SNR signals, [27] proposes wavelet-based digital signal processing algorithms to encounter the high power non-stationary noise in infrared wireless systems.

### 2.1.2 Empirical Mode Decomposition Based Denoising

Proposed by Huang et al [28], empirical mode decomposition (EMD) operates in an iterative process to generate several components of the original signal, which for a signal $f(t)$ can be defined as in [29] –

$$f(t) = \sum_{j=1}^{M} \emptyset_j(t) \qquad (4)$$

Where $\emptyset_j(t)$, the mono-component signals with amplitude $r(t)$ are called the Intrinsic Mode Function (IMF), which characterizes the intrinsic and reality information of the decomposed signal. The process by which EMD functions can be described well by an algorithm than mathematical theories as can be seen in [30 - 31]. To elaborate, EMD is an adaptive process which decomposes a multicomponent signals into several IMFs as mentioned previously [32]. In order to create the IMFs sifting process is employed where cubic spline interpolation locates the local maxima and minima to form an upper envelope and lower envelope. Subtracting the mean of these two envelopes from the original signal results in the formation of IMFs with certain characteristic properties. As the multicomponent signals are decomposed to several IMFs, denoising requires identifying the noise components so that they can be removed and the original signal can be reconstructed without noise contribution. In [29], the reconstruction is described as the process where the IMFs with useful information are combined along with some residual noise $r[n]$ and can be defined as –

$$x(n) = \sum_{i=1}^{n} IMF[n] + r[n] \qquad (5)$$

Research in [29] refers to an implementation of EMD block in GNU Radio [33], an open source software that hosts signal processing packages [34]. Experimental setup in [29] aims to reduce noise contribution in received signals and improve the transmission bit error rate (BER).

One prominent performance factor for the sifting process is the right estimation of when to stop the sifting process. Besides the stopping criterion determination right method of spline interpolation is also necessary to generate desired results from the EMD. Different spline interpolation methods are tried in [29] and results were compared to analyze the accuracy of the sifting process for EMD. Similar to [29], authors in [31] pointed out the possibility of erroneous outputs of EMD because of the convergence problems in sifting process and the correct choice of interpolation methods.

Several other research works focused on the issue of implementing the right method of spline interpolation for the sifting process. In [35], an alternative to cubic spline interpolation, B-spline is introduced with no significant improvements. Iterating filters are considered in [36] to resolve the issue of a convergence problem. IMFs are analyzed based on their energy difference and is considered to be useful for differentiating purposes in [32]. With the replacement of cubic spline with a rational spline, work in [37] presents some promising results. In [38], authors present IMF threshold determination based denoising technique inspired by the threshold determination technique in wavelet-based denoising. Coherent with the threshold selection mechanism in wavelet, work in [38] suggests the use of the same principles with the only difference of applying the threshold to each sample of every IMF instead of applying the threshold to only reconstructed signals, which is the case in wavelet.

## 2.2 Matrix Factorization Based Techniques

Matrix factorization employs a mathematical approach to decompose a matrix and has been used in many applications to solve numerous problems [39]. In the context of noise cancellation during signal processing, matrix factorization technique has been put to use as it enables signal space analysis. Two of the matrix factorization techniques, singular value decomposition, and non-negative matrix factorization are discussed in the following sections.

### 2.2.1 Singular Value Decomposition Based Denoising

Singular Value Decomposition (SVD) is one of the useful matrix decomposition methods that enables the factorization of a matrix. For a matrix **A**, SVD factorizes **A**n into the product of a unitary matrix **U**, a diagonal matrix **Σ**, and another unitary matrix $\mathbf{V}^H$ [39]. If matrix **A** is $m \times n$ matrix, **U** will be the unitary matrix $m \times m$, with non-negative real numbers $m \times n$ diagonal matrix is **Σ** and $\mathbf{V}^H$ is the $n \times n$ unitary matrix. The $\Sigma_{i,j}$ of **Σ** are the singular values of **A** and the left-singular vectors of **A** are the $m$ columns of **U** while right-singular vectors are the $n$ columns of **V**. SVD being numerically stable produces non-negative eigenvalues which makes it a preferable choice over Eigen decomposition [39].

As discussed in the previous sections, signal processing to sense the availability of spectrum is one of the primary tasks of cognitive radio. Second-order statistical data and covariance matrix are commonly used methods to analyze a set of data, in this case, which would be sensed spectrum. The above-mentioned matrix factorization technique opts to reduce the dimension of the sample covariance matrix retaining an important set of information which can be used to distinguish different components of a signal such as noise. In [9], the factorization technique – SVD is employed to detect noise anomalies in the 2.4 GHz band. It is notable to point out that SVD was employed instead of Eigen decomposition to differentiate the noise components in the signal. SVD's numerical stability and non-negative eigenvalue output make it a desirable choice for data analysis. Work in [9] first defines an unbiased data matrix, **M** in order to decompose the sample covariance matrix of the sensed 2.4GHz band and is expressed as -

$$\mathbf{M} = \frac{1}{\sqrt{N_f - 1}}(\mathbf{X} - E[\mathbf{X}]) \qquad (6)$$

Then SVD of matrix M results in the factorization, as defined in (6) –

$$\mathbf{M} = \mathbf{W}\Sigma\mathbf{U}^H \qquad (7)$$

The $\Sigma_{i,j}$ of **Σ** are the singular values of **M** and from the knowledge of principal component analysis (PCA), the columns of **W** that are known to be the left singular vectors are the eigenvectors of $\mathbf{MM}^H$ whereas the columns of **U** which are the right singular vectors are known to be the eigenvectors of $\mathbf{M}^H\mathbf{M}$. From the decomposed covariance matrix we can get the eigenvectors and eigenvalues, as described by [40 - 42]. With these obtained eigenvectors [9] creates $\widetilde{X}$ of the measured data *X*, a projection of the measured data that only contains the strongest signal space component and is expressed as –

$$\widetilde{X} = \sum_{i=1}^{L} w_i w_i^H (X - E[X] + E[X]) \qquad (8)$$

The *L* components refer to the L principal components of X. Because of the necessity to perform, PCA requires mean subtraction which results in adding $E[\mathbf{X}]$ to **X**. This addition results in generating undesired outputs from SVD and disrupts the non-negative constraints.

### 2.2.2 Non-Negative Matrix Factorization Based Denoising

Non-negative factorization (NMF), also referred to as non-negative approximation, of matrix results in non-negative outputs which makes it easier to analyze the signal of interest [43-44]. NMF factorizes a matrix **A** into two matrices **W** and **H**, all of the three with a common property of having no negative elements. To elaborate, a matrix **A** made up of $m \times n$ matrix can be factorized into **W**, a $m \times p$ matrix and **H** as $n \times p$ matrix where $p$ can be significantly lower than both $m$ and $n$. **H** is the coefficient matrix that supplies with appropriate coefficients for the numerical approximation NMF provides with its multivariate analytical characteristics. To track the divergence of the factorized matrix **A** and the product matrix **W, H** different divergence function, also referred to as cost functions, can be defined for the purpose of introducing regulations.

Keeping in mind this case-specific problem of SVD, authors in [9] employed NMF as the second technique for denoising purpose. Dimension reduction technique like NMF allows the creation of two non-negative matrices as outlined earlier in the section. NMF algorithm factorizes the output **W,** which is the feature matrix, and **H,** the coefficient matrix followed by a low-rank approximation, to estimate data matrix **X** through –

$$X = \sum_{j=1}^{L} u_j w_j^H + R \qquad (9)$$

Once again, rank $L$ is the number of principal components. For the process of factorization to continue and produce the desired result a divergence function is defined that regulates the difference between the data matrix X and **UW**. Such function in [9]] measures the difference by defining a cost/divergence function D(X, UW) and uses Kullback-Leiber (KL) as a cost function to confirm convergence [45-49].

Both the techniques are suitable for achieving noise cancellation by providing means to clearly identify the signal space from the noise space. In short, SVD and NMF are able to decompose the unprocessed signal to capture principal independent components which in turn spaces out the signal components making the dataset convenient to inspect. It is also pointed out in [9] that certain parameter adjustment is required, especially the value for $L$ principal components in the factorization techniques. A clear overview of the importance of non-stationary noise removal in the context of cognitive radio is highlighted at the beginning of [9]. A performance evaluation along with the methodology to setup noise removal experiment is also discussed in [9].

## 2.3 Adaptive Filter Based Techniques

Adaptive filters based denoising technique requires filter design that can embrace the randomness of noise and operate by readjusting the filter parameters in a recursive manner to perform noise cancellation. Precisely, an input signal $x(n)$, to the adaptive filter is updated with a weight coefficient $w(n)$ to produce the output signal, $y(n)$ expressed as –

$$y(n) = w(n).x(n) \qquad (10)$$

Starting the filter operation with a randomly selected weight factor, adaptive filters have to rely on a feedback mechanism to minimize the residual noise present in the noisy signal. An adaptive filter in Figure 2, shows an approach to calculate the output signal [50] [51]. The difference between desired signal $d(n)$ and updated signal $y(n)$ is considered to be the feedback or error signal $e(n)$, written as –

$$e(n) = d(n) - y(n) \qquad (11)$$

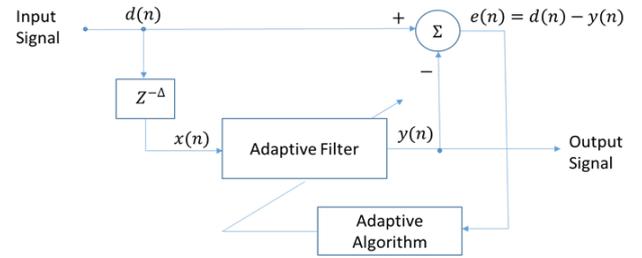

Fig. 2. Block diagram representation of feedback ANC system

Adaptive algorithms in charge of readjusting the filter parameters can be grouped as gradient-based and non-gradient based algorithms [52-53]. Gradient descent is a multivariate optimization technique that starts with an assigned initial value and follows the negative of the gradient to reach the desired local minimum. At the initial stage of operation, all gradient-based algorithms employ a *step size,* which can be described as the guiding factor to decide on the direction of the negative descent from one point to the next. It is preferred that the *step size* is chosen small so as to achieve the optimal convergence speed while maintaining a small Steady State Misadjustment (SSM) for stable optimization.

### 2.3.1 Least Mean Square Algorithm Based Denoising

Mostly known for ease of implementation and computational efficiency, one of the gradient-based algorithms that prevailed the test of time is Least Mean Square (LMS) [53]. In LMS algorithms weight coefficients $w(n)$ are adjusted with a predefined *step size* μ and expressed as,

$$W(n + 1) = W(n) + \mu\, e(n)X(n) \qquad (12)$$

Although LMS algorithm has a fast convergence rate, fixed *step size* calculation severely degrades the performance of LMS and calls for improvement [50].

### 2.3.2 Normalized Least Mean Square Algorithm Based Denoising

Normalized LMS (NLMS) is an extension of the conventional LMS algorithm. In NLMS, weight coefficients are updated with a revised *step size*, resulting in the following –

$$W(n+1) = W(n) + \frac{\mu}{\gamma + X^T(n)X(n)} e(n)X(n) \quad (13)$$

Where, $\gamma > 0$ and $0 < \mu < 1$

NLMS delivers faster convergence rate compared to LMS but still is far from achieving the optimum tradeoff between convergence rate & SSM (Steady State Misadjustment). However, based on simplicity and inexpensive computational requirements, both LMS and NLMS has been widely used in adaptive noise cancellation schemes [52] [54].

Acknowledging the simplicity in implementation and satisfying performance, [10]] [55 - 56] studies use of Least Mean Square (LMS) and Normalized LMS in a cognitive radio system built on software defined radio. Contrary to the popular practice of simulated implementations, authors in [10] intended to design a practical setup of NLMS based channel equalizers and conduct a real-time measurement of transmission error rate over noise power (*BER vs Eb/No*). Based on the results from simulations in [55 - 56], authors in [10] points out LMS based equalizer is inefficient and extends their work to implement NLMS based channel equalizer in a practical setup, based on several other research work [57 - 58]. Results from the experiments in [10] suggest channel equalizers with NLMS algorithm is better in terms of simplicity and performance when compared to a traditional LMS algorithm. In [59], research work is focused on denoising process of the sensed spectrum. Spectrum sensing techniques based on energy detection employ a threshold value against which sensed signals are compared to evaluate the availability of signals.

## 3 Performance Comparison of Denoising Techniques for Cognitive Radio

The three categories of denoising techniques discussed in this paper are analyzed in terms of their strengths and weaknesses, as shown in Table 1. Time-frequency analysis based approach decomposes the received signal into many levels and allows inspection in both time-frequency domains. Whereas, matrix factorization techniques decompose a sparse or large matrix data set into significantly smaller dimensions, projecting out only the strong signal components. In contrast to the first two techniques, adaptive filter based denoising makes use of traditional filtering process with adaptive algorithms that work in a recursive fashion to cancel out the noise.

### 3.1 Performance of time-frequency analysis based techniques

Wavelet transform based denoising can locate multiple edges or singularities of the signal under analysis. However, wavelet transform based denoising requires a fixed basis function, a chosen mother wavelet around which revolves the process of wavelet analysis of a received signal. In Empirical Mode Decomposition (EMD), received signal is dissected into several mono-component signals called the Intrinsic Mode Function using sifting algorithm. Defining the appropriate stopping criterion for sifting algorithm and choosing the right spline interpolation method is essential in achieving desired output from EMD based denoising [29].

### 3.2 Performance of matrix factorization based techniques

Singular value decomposition (SVD), in the context of denoising a signal, factorizes the sample covariance matrix of the signal into sample eigenvectors and eigenvalues [9]. SVD is numerically stable but is data-driven as it needs to obtain data from the signal. Non-negative factorization (NMF), just like SVD is also reduced the original data and can produce non-negative outputs. NMF constructs a low-rank approximation of the original data matrix. Proper constraints for the low-rank approximation is, therefore, a key factor that needs to be determined in initial stages of NMF algorithms [9].

### 3.3 Performance of adaptive filter based techniques

LMS algorithm requires an optimal *step size* to be defined at the initial stage of the adaptive filtering process. Selecting the right *step-size* is a crucial factor for any gradient descent based adaptive algorithms like LMS. Proper *step-size* estimation leads to achieve an optimal trade-off between convergence rate and stability. As mentioned in the earlier section for adaptive filters, NLMS is an extension of the LMS algorithm. Improvements in

NLMS algorithm is mostly the revised *step size* that includes more criterion to estimate an optimal *step size*. Both the adaptive filter based techniques are easy to implement and consumes less computational resources.

Table 1. Denoising techniques outlined in terms of strengths and weaknesses.

| | Wavelet Transform | |
|---|---|---|
| | Strength | Weakness |
| Time-Frequency Analysis Based Denoising | • Allows non-linear signal analysis.<br>• Locates multiple edges and singularities in the signal. | • Fixed basis function<br>• The threshold can only be applied to each level of decomposition. |
| | Empirical Mode Decomposition | |
| | Strength | Weakness |
| | • Adaptive.<br>• Threshold can be applied to every sample of each IMF. | • Choosing right stopping criterion.<br>• Choosing the right spline-interpolation method |
| Matrix Factorization Based Denoising | Singular Value Decomposition | |
| | Strength | Weakness |
| | • Numerically stable<br>• Produce non-negative outputs. | Data-driven - Needs data from the sample covariance matrix. |
| | Non-negative Matrix Factorization | |
| | Strength | Weakness |
| | Non-negative outputs | Cannot guarantee convergence to a global optimum solution |
| Adaptive Filter Based Denoising | Least Mean Square Algorithm | |
| | Strength | Weakness |
| | • Easy to implement<br>• Computationally simple. | Difficult to achieve optimum trade-off between convergence rate and stability |
| Adaptive Filter Based Denoising | Normalized LMS | |
| | Strength | Weakness |
| | Faster convergence rate | Yet to achieve better trade-off between convergence rate and stability |

## 4 Conclusions

To reiterate, we provide a review of denoising techniques that can be implemented for cognitive radio. These denoising techniques are categorized as 1) Time-frequency analysis, 2) Matrix factorization and 3) Adaptive filter based denoising techniques. Time-frequency analysis based techniques such as empirical mode decomposition and wavelet transform based denoising are discussed and analyzed. An overview of related research works, where the denoising techniques were applied is also provided. Matrix factorization techniques that are discussed in this paper are singular value decomposition and non-negative matrix factorization. A detailed analysis of both the techniques are provided and their effectiveness in the context of denoising non-stationary signals are reviewed. Conventional filter based noise cancellation with adaptive filters are also reviewed but the discussion on this technique is provided in the context of using adaptive filters for denoising in cognitive radio. Least Mean Square (LMS) and Normalized LMS are the two adaptive filter based denoising techniques analyzed in this paper. Implementation and drawbacks of these two techniques are studied by discussing some of the related research works where adaptive filter based denoising is applied on the cognitive radio system. Finally, to compare the performance of the reviewed techniques, a tabular presentation of their strengths and weaknesses are outlined at the end.

*References:*
[1] Federal Communications Commission, "Spectrum Policy Task Force," Rep. ET Docket no. 02-135, Nov. 2002.
[2] Mohsen RiahiManesh, Adnan Quadri, Sriram Subramanian, and Naima Kaabouch, "An Optimized SNR Estimation Technique Using Particle Swarm Optimization Algorithm," *IEEE Consumer Communications and Networking Conference*, pp. 1-7, 2017.
[3] M. McHenry, "Frequency agile spectrum access technologies," *FCC Workshop Cogn. Radio*, May 19, 2003.
[4] Adnan Quadri, Mohsen RiahiManesh, and Naima Kaabouch, "Denoising Signals in Cognitive Radio Systems: An Evolutionary Algorithm Based Adaptive Filter," *IEEE Annual Ubiquitous Computing, Electronics & Mobile Communication Conference*, pp. 1-6, March 18 2016.
[5] Manesh, Mohsen Riahi, and Naima Kaabouch. "Interference Modeling in Cognitive Radio Networks: A Survey." arXiv preprint arXiv:1707.09391 (2017).
[6] Manesh, Mohsen Riahi, Naima Kaabouch, Hector Reyes, and Wen-Chen Hu. "A Bayesian model of the aggregate interference power in cognitive radio networks." In *Ubiquitous Computing, Electronics & Mobile Communication Conference, IEEE Annual*, pp. 1-7. IEEE, 2016.
[7] Riahi Manesh, Mohsen, Naima Kaabouch, Hector Reyes, and Wen-Chen Hu. "A Bayesian approach to estimate and model SINR in wireless networks." International Journal of Communication Systems 30, no. 9 (2017).
[8] X. Li, F. Hu, H. Zhang and C. Shi, "Two-branch wavelet denoising for accurate spectrum sensing in cognitive radios," *Telecommunication Systems*, vol. 57, no. 1, pp. 81-90, 2013.
[9] J. van Bloem, R. Schiphorst and C. Slump, 'Removing non-stationary noise in spectrum sensing using matrix factorization', *EURASIP Journal on Advances in Signal Processing,* vol. 2013, no. 1, p. 72, 2013.
[10] R. Martinek and J. Zidek, "The Real Implementation of NLMS Channel Equalizer into the System of Software Defined Radio," *AEEE*, vol. 10, no. 5, 2012.
[11] A. Das, B. Chatterjee, S. Pattanayak and M. Ojha, 'An Improved Energy Detector for Spectrum Sensing in Cognitive Radio System with Adaptive Noise Cancellation and Adaptive Threshold', *Computational Advancement in Communication Circuits and Systems*, pp. 113-119, 2015.
[12] A. Sonnenschein and P. M. Fishman, "Radiometric detection of spread spectrum signals in noise of uncertainty power," *IEEE Transactions on Aerospace and Electronic Systems*, vol. 28, no. 3, pp. 654–660, 1992.
[13] Adnan Quadri, Mohsen RiahiManesh, and Naima Kaabouch, "Performance Comparison of Evolutionary Algorithms for Noise Cancellation in Cognitive Radio Systems," *IEEE Consumer Communications and Networking Conference*, pp.1-6, 2017..
[14] R. Tandra and A. Sahai, "Fundamental limits on detection in low SNR under noise uncertainty," *Proceedings of the International Conference on Wireless Networks, Communications and Mobile Computing*, vol. 1, pp. 464–469, Maui, Hawaii, USA, June 2005.
[15] Y. Zeng, Y. Liang, A. Hoang and R. Zhang, "A Review on Spectrum Sensing for Cognitive Radio: Challenges and Solutions," *EURASIP Journal on Advances in Signal Processing*, vol. 2010, pp. 1-16, 2010.
[16] Kaabouch, Naima, ed. Handbook of Research on Software-Defined and Cognitive Radio Technologies for Dynamic Spectrum Management. IGI Global, 2014.
[17] V. Gupta and A. Kumar, "Wavelet Based Dynamic Spectrum Sensing for Cognitive Radio under Noisy Environment," *Procedia Engineering*, vol. 38, pp. 3228-3234, 2012.
[18] F. Luisier, T. Blu, and M. Unser, "A new SURE approach to image denoising: Inter-scale orthonormal wavelet thresholding," *IEEE Trans. Image Processing*, vol. 16, no. 3, pp. 593–606, Mar. 2007.
[19] I. Daubechies, Ten Lectures on Wavelets. Philadelphia: SIAM, 1992.
[20] C.S. Burrus, R.A. Gopinath, and H. Guo, Introduction to Wavelets and Wavelet Transforms, a Primer. Upper Saddle River, NJ (USA): Prentice Hall, 1998.
[21] K. Divakaran, N. P and S. R. R, "Wavelet Based Spectrum Sensing Techniques for Cognitive Radio - A Survey," *International Journal of Computer Science and Information Technology*, vol. 3, no. 2, pp. 123-137, 2011.

[22] H. Wang, Y. Xu, X. Su and J. Wang, "Cooperative Spectrum Sensing with Wavelet Denoising in Cognitive Radio," *IEEE 71st Vehicular Technology Conference*, 2010.

[23] Z. Tian and G. Giannakis, "A Wavelet Approach to Wideband Spectrum Sensing for Cognitive Radios," *1st International Conference on Cognitive Radio Oriented Wireless Networks and Communications*, 2006.

[24] M.H.C. Dias and G.L. Siqueira, "On the Use of Wavelet-Based Denoising to Improve Power Delay Profile Estimates from 1.8 GHz indoor wideband measurements," *Wireless Personal Communications*, Vol. 32, No. 2, pp. 153–175, January 2005.

[25] S.D. Mantis, "Localization of Wireless Communication Emitters Using Time Difference of Arrival (TDOA) Methods in Noisy Channels," *Master's Thesis, Naval Post Graduate School*, Monterey, CA. 2001.

[26] R.S.Wong and V.K. Bhargava, "Denoising of low SNR Signals Using Composite Wavelet Shrinkage," *IEEE Pacific Rim Conference on Communications, Computers and Signal Processing*, vol. 1, pp. 302–305, August 1997.

[27] X. Fernando, S. Krishnan, and H. Sun, "Adaptive Denoising at Infrared Wireless Receivers," *17th Annual Aerosense Symposium, Florida*, Orlando, April 2003.

[28] N. Huang, Z. Shen, S. Long, M. Wu, H. Shih, Q. Zheng, N. Yen, C. Tung and H. Liu, "The empirical mode decomposition and the Hilbert spectrum for nonlinear and non-stationary time series analysis," *Proceedings of the Royal Society A: Mathematical, Physical and Engineering Sciences*, vol. 454, no. 1971, pp. 903-995, 1998.

[29] S. Mankad and S. Pradhan, "Application of Software Defined Radio for Noise Reduction Using Empirical Mode Decomposition," *Advances in Intelligent and Soft Computing*, pp. 113-121, 2012.

[30] P. Flandrin, G. Rilling, and P. Goncalves, "Empirical mode decomposition as a filter bank," *IEEE Signal Processing Letters*, vol. 11, pp. 112–114, Feb. 2004.

[31] G. Rilling and P. Flandrin, "One or two frequencies the empirical mode decomposition answers," *IEEE Trans. Signal Processing*, pp. 85–95, Jan. 2008.

[32] C. Junsheng, Y. Dejie and Y. Yu, "Research on the intrinsic mode function (IMF) criterion in EMD method," *Mechanical Systems and Signal Processing*, vol. 20, no. 4, pp. 817-824, 2006.

[33] Gnuradio.org, 2015. [Online]. Available: http://www.gnuradio.org.

[34] Bard J., Kovarik V.J., Software Defined Radio- The Software Communications Architecture. John Willey and Sons. (2007)

[35] Chen Q., Huang, N., Riemenschneider S., and Xu Y., "A B-spline approach for Empirical Mode Decompositions." *Advances in Computational Mathematics*, vol. 24, pp. 171–195, 2006

[36] L. Lin, Y. Wang and H. Zhou, "Iterative Filtering As An Alternative Algorithm For Empirical Mode Decomposition," *Advances in Adaptive Data Analysis*, vol. 01, no. 04, pp. 543-560, 2009.

[37] M. Peel, G. Pegram and T. Mcmahon, "Empirical mode decomposition: Improvement and application," ResearchGate, 2007. [Online]. Available:https://www.researchgate.net/publication/229022029_Empirical_mode_decomposition_Improvement_and_application. [Accessed: 11- Dec- 2015].

[38] Y. Kopsinis and S. McLaughlin, 'Empirical Mode Decompisition Bsed Denoising Techniques', *IAPR Workshop on Cognitive Information Processing,* 2008.

[39] G. Stewart, "On the Early History of the Singular Value Decomposition", *SIAM Rev.*, vol. 35, no. 4, pp. 551-566, 1993.

[40] TW Anderson, An Introduction to Multivariate Statistical Analysis, 2nd edn. Wiley series in probability and mathematical statistics, John Wiley & Sons, New York, 1984.

[41] F Castells, P Laguna, L Sörnmo, A Bollmann, JM Roig, "Principal component analysis in ECG signal processing," *EURASIP J. Appl. Signal Process*. 2007, pp. 98–98, 2007.

[42] S Aviyente, EM Bernat, SM Malone, WG. Iacono, "Time-frequency data reduction for event related potentials: combining principal component analysis and matching pursuit." *EURASIP J. Adv. Signal Processing*, pp. 1–13, 2010.

[43] Inderjit S. Dhillon, SuvritSra, "Generalized Nonnegative Matrix Approximations with

Bregman Divergences," *Neural Information Processing Systems Conference*, 2005. [Online]. Available:http://papers.nips.cc/paper/2757-generalized-nonnegative-matrix-approximations-with-bregman-divergences.pdf

[44] TandonRashish, SuvritSra, "Sparse nonnegative matrix approximation: new formulations and algorithms", *Max Planck Institute for Biological Cybernetics*, TR, 2010

[45] F. Weninger, B. Schuller, A. Batliner, S. Steidl, D. Seppi, "Recognition of nonprototypical emotions in reverberated and noisy speech by nonnegative matrix factorization," *EURASIP J. Adv. Sig. Proc.*, pp. 1–16, 2011.

[46] DD. Lee, HS. Seung, "Algorithms for non-negative matrix factorization," *Advances in Neural Information Processing*, vol. 13, pp. 556–562, MIT Press, 2000.

[47] MW. Berry, M. Browne, AN. Langville, VP. Pauca, RJ. Plemmons, "Algorithms and applications for approximate nonnegative matrix factorization," *Computational Statistics and Data Analysis*, vol. 52, pp. 155–173, 2006.

[48] Z. Yang, H. Zhang, Z. Yuan, E Oja, "Kullback-Leibler divergence for nonnegative matrix factorization," *Artificial Neural Networks and Machine Learning—ICANN 2011*, vol. 6791 of Lecture Notes in Computer Science, pp. 250–257, 2011.

[49] O. Okun, H. Priisalu, "Fast nonnegative matrix factorization and its application for protein fold recognition," *EURASIP J. Appl. Signal Process*, pp. 649–656, 2006.

[50] R. Ramli, A. Abid Noor and S. Abdul Samad, "A Review of Adaptive Line Enhancers for Noise Cancellation', *Australian Journal of Basic and Applied Sciences*, vol. 6, no. 6, p. 337-352, 2012.

[51] Widrow and S. Stearns, Adaptive signal processing. Englewood Cliffs, N.J.: Prentice-Hall, 1985.

[52] N. George and G. Panda, 'Advances in active noise control: A survey, with emphasis on recent nonlinear techniques', *Signal Processing,* vol. 93, no. 2, pp. 363-377, 2013.

[53] S. Haykin, Adaptive filter theory. Third Edition, New York: Prentice-Hall, 2002.

[54] Siddappaji and K. Sudha, 'Performance analysis of New Time Varying LMS (NTVLMS) adaptive filtering algorithm in noise cancellation system', 2015 *International Conference on Communication, Information & Computing Technology*, 2015.

[55] Alok Pandey, L.D. Malviya, Vineet Sharma, "Comparative Study of LMS and NLMS Algorithms inAdaptive Equalizer," *International Journal of Engineering Research and Applications,* vol. 2, Issue 3, pp.1584-1587, May-Jun 2012.

[56] Anita Garhwal, ParthaPratim Bhattacharya, "Performance Enhancement of WiMAX System using Adaptive Equalizer," *Journal of Emerging Trends in Computing and Information Sciences*, vol. 3, Issue no. 4, April 2012.

[57] M. Radek, M. Al. Wohaishi, J. Zidek, "Software Based Flexible Measuring Systems for Analysis of Digitally Modulated Systems," *The 9th Roedunet International conference*, pp. 397 -402, IEEE 2010.

[58] M. Radek, J. Zidek, K. Tamela and L. Klein, "Implementation of LMS Equalizer into Software Defined Radio System SDR," *13th International Conference on research in Telecommunication Technologies*, pp. 75-79, 2011.

[59] D. Cabric, S. Mishra, R. Brodersen, "Implementation issues in spectrum sensing for cognitive radios," *Proceedings Asilomar Conference on Signals, Systems and Computers*, vol. 1, pp. 772–776, 2004.